\documentstyle[12pt,epsf]{article}
\newcommand{\ra}{\rangle}
\newcommand{\vac}{|0\rangle}
\newcommand{\la}{\langle}

\newcommand{\eps}{\epsilon}

\newcommand{\OO}{{\cal O}}

\newcommand{\be}{\begin{equation}}
\newcommand{\ee}{\end{equation}}
\newcommand{\ben}{\begin{eqnarray}\displaystyle}
\newcommand{\een}{\end{eqnarray}}
\newcommand{\refb}[1]{(\ref{#1})}

\begin{document}

{}~ \hfill\vbox{\hbox{hep-th/9809111}\hbox{MRI-PHY/P980960}
}\break

\vskip 3.5cm

\centerline{\large \bf Type I D-particle and its Interactions}

\vspace*{6.0ex}

\centerline{\large \rm Ashoke Sen
\footnote{E-mail: sen@mri.ernet.in, asen@thwgs.cern.ch}}

\vspace*{1.5ex}

\centerline{\large \it Mehta Research Institute of Mathematics}
 \centerline{\large \it and Mathematical Physics}

\centerline{\large \it  Chhatnag Road, Jhoosi,
Allahabad 211019, INDIA}

\vspace*{4.5ex}

\centerline {\bf Abstract}

In a previous paper (hep-th/9808141) we showed that the type I
string theory contains a stable non-BPS D-particle carrying SO(32)
spinor charge.  In this paper
we formulate the rules for computing the spectrum and interaction
of open strings with one or both ends lying on this D-particle. 

\vfill \eject

\baselineskip=18pt

The discovery that D-branes carry Ramond-Ramond (RR)
charge\cite{POLC} $-$ and hence
represent stable BPS saturated particles in string
theory $-$ has dramatically changed our understanding of string
theory. The inclusion of (wrapped) D-branes in the
particle spectrum has provided key tests of many duality conjectures.
Most of these tests have focussed on BPS particles in
supersymmetric string theories; although there are some notable
exceptions where non-BPS D-branes in non-supersymmetric string
theories have been used to test duality conjectures involving
non-supersymmetric string theories\cite{NONSUSY}.

Although comparing the spectrum of BPS states in two theories
provides us with a stringent test of duality between these two
theories, quite often non-BPS states may also be useful in
testing a duality conjecture. Thus for example if one of the
theories contains, in its perturbative spectrum, a non-BPS
state which is
stable due to charge conservation (being the state of minimum
mass carrying a given charge quantum number), then the dual
theory must also contain a stable non-BPS state in its spectrum
carrying the same quantum numbers. Several examples of this kind
were discussed in \cite{NONBPS}. 
One such example involved the
duality relating the SO(32) heterotic and the type I string
theories. The heterotic theory contains massive non-BPS states in
the spinor representation of SO(32) which must be stable due to
charge conservation. Thus the type I theory must also contain states
in the spinor representation of SO(32). It was suggested in
\cite{NONBPS} that these states correspond 
to stable non-BPS D-particles in type I, although no 
explicit construction of such D-particles was known at that time.

This question was addressed in \cite{SPINOR} from a slightly
different angle. The starting point in \cite{SPINOR} was a
D-string anti- D-string pair of type I string theory. There is a
tachyon field $T$ living on the world volume of this system. The
tachyon potential $V(T)$ is invariant under $T\to -T$ and hence
has a doubly degenerate minimum at $\pm T_0$. Thus one can
construct a kink configuration on the world volume of this system
which interpolates between the tachyonic vacua $\pm T_0$. It was
shown in \cite{SPINOR} that this kink solution carries SO(32)
spinor charge and has finite mass, and therefore represents
the state that we have been searching for. Apparently this
description is very different from that of a D-particle in type I.
However it was found that this kink configuration admits a
description in terms of a solvable boundary conformal field
theory; and in this description it indeed corresponds to a
D-particle for which ends of open strings satisfy Dirichlet
boundary condition in all nine spatial directions.

Regarding the D-particle of type I as the tachyonic kink solution
on the D-string anti- D-string pair, and
following the algorithm of ref.\cite{SPINOR}, we can in principle
derive the
complete set of rules for the spectrum and interaction of open
strings whose one or both ends lie on the D-particle. However,
in this paper we shall take a slightly different approach.
Instead of trying to derive the interaction rules from
\cite{SPINOR} directly, we shall try 
to formulate a set of rules without
any reference to the kink solution,
take them to be the defining relations for a D-particle
of type I and check the internal consistency of these rules. Of
course in formulating the rules we shall be using the insight
gained from \cite{SPINOR}.  

As pointed out in \cite{SPINOR}, this D-particle can also
be regarded as a configuration of type IIB string theory,
although it has a tachyonic mode and hence is unstable.
We shall find it more convenient to
first describe the rules in the context of type IIB
string theory, and specify the
action of the world-sheet parity operator on the various open
string modes. The open string spectrum in type I string theory is
then obtained by projecting onto $\Omega$-invariant states
(vertex operators).
The interaction rules involving open strings ending on
these D-particles can be obtained by following
the usual procedure of including unoriented world-sheet diagrams
and extra combinatoric factors.

The D-particle of type IIB string theory, constructed this way,
is in fact identical to the non-BPS D-particle discussed in
\cite{BERGAB}.\footnote{This was in fact originally motivated by
an attempt to construct a D-particle state in type I string
theory\cite{PRIV}.} This paper dealt with type IIB string theory in
the presence of an orbifold plane with a D-particle
stuck to the orbifold plane, and constructed the boundary state
describing the D-particle\footnote{This state also has an
alternate but equivalent description as a soliton on the D-string
anti- D-string pair\cite{BOUND}.}. The tachyon in the spectrum was
removed by the orbifold projection instead of $\Omega$
projection as in the present case. However, since the boundary state
describing a D-brane does not contain full information about the
Chan Paton factors, it was not obvious that the D-particle
described in this manner admits a consistent set of interaction
rules. The present paper shows that it is indeed possible to
construct such interaction rules.

We shall now specify the rules for computing the spectrum and
interactions of open strings which end on the D-particle.
Since besides the non-BPS
D-particle, the type IIB string theory also contains BPS D$p$-branes
for odd $p$, we need to specify the rules for the spectrum and
interaction of open strings in the 0-0, 0-$p$ and $p$-0 sector.

\noindent{\bf 0-0 sector}: In this sector we have
Dirichlet-Dirichlet (DD) boundary condition on all the space-like
coordinates $X^1,\ldots X^9$ and Neumann-Neumann (NN) boundary
condition on the time coordinate $X^0$. There is no GSO
projection on these open string states so that
before the $\Omega$
projection the open string spectrum includes all the Fock space
states. We associate a $2\times 2$ Chan-Paton factor
with each vertex operator in this sector, with the restriction
that for states which are even under $(-1)^F$, the Chan Paton
factor is the identity matrix, whereas for states which are
odd under $(-1)^F$, the Chan Paton factor is given by
\be \label{en1}
\sigma_2 = \pmatrix{0 & -i\cr i & 0}\, .
\ee
As usual, we take the Fock vacuum to be odd under $(-1)^F$.

In order to find the spectrum of 0-0 open strings in the type I
theory we need to know the action of $\Omega$ on different open
string states.
$\Omega$ acts by transposition on the Chan Paton factors and has
the usual action (see, for example \cite{GIMPOL})
on the various oscillators:
\be \label{en1a}
\alpha^\mu_r\to \pm e^{i\pi r}\alpha^\mu_r, \qquad
\psi^\mu_r\to \pm e^{i\pi r}\psi^\mu_r\, ,
\ee
where the upper (lower) sign holds for NN (DD) boundary
condition, and $\alpha^\mu_r$ and $\psi^\mu_r$ denote the
oscillators associated with the bosonic and fermionic fields
$X^\mu$ and $\psi^\mu$ respectively. Finally, the action of $\Omega$ on
the Fock vacuum $\vac$ is taken to be identity in computing its
action on  $(-1)^F$ odd states, and $-i$ for computing its action
on the $(-1)^F$ even states. Thus for example,
according to these rules, the Fock space states:
\be \label{en2}
\vac, \qquad \hbox{and} \qquad \psi^m_{-{1\over 2}}\vac, \quad
\hbox{for}\quad 1\le m\le 9\, ,
\ee
are $\Omega$ even. However the state $\vac$, being $(-1)^F$ odd,
has a Chan Paton factor $\sigma_2$ associated with it. Since the
action of $\Omega$ changes the sign of $\sigma_2$, the state
corresponding to $\vac$ is actually odd under $\Omega$ and hence
is projected out in the type I theory. 
This rule can be simplified by ignoring the
action of $\Omega$ on the Chan Paton factors, and taking the
$\Omega$ charge of the Fock vacuum to be $-1$ and $-i$ for
$(-1)^F$ odd and even states respectively.\footnote{Although
these rules seem completely ad hoc, they are the ones which
emerge out of the analysis of ref.\cite{SPINOR}. Also, as we
shall show, they are internally consistent.}

\noindent{\bf 0-$p$ and $p$-0 sectors}: Let us take the
D$p$-brane to extend along $X^1,\ldots X^p$ direction. Then in
the Neveu-Schwarz (NS) sector, the
oscillators associated with the fermionic fields
$\psi^1,\ldots\psi^p$ are integer moded and the oscillators
associated with the fermionic fields
$\psi^0,\psi^{p+1},\ldots\psi^9$ are half integer moded. In the R
sector the situation is reversed, $-$
the oscillators associated with the fields
$\psi^1,\ldots\psi^p$ are half integer moded and the oscillators
associated with the fields
$\psi^0,\psi^{p+1},\ldots\psi^9$ are integer moded. The allowed
vertex operators (including the Chan Paton factors) in the 0-$p$
sector are taken to be of the form:
\be \label{e1}
\pmatrix{1\cr 0} \otimes V_s + i\eps\pmatrix{0\cr 1} 
\otimes V_{s'}\, ,
\ee
where $V_s$ is a $(-1)^F$ even vertex operator, and $V_{s'}$ is a
$(-1)^F$ odd vertex operator, related to $V_s$ through the
relation:
\be \label{e3}
|s'\ra = U|s\ra\, ,
\ee
where,
\be \label{e3a}
U = (-1)^{F'} \psi_0^1\ldots \psi_0^p \, ,
\ee
in the NS sector, and
\be \label{e3b}
U = i (-1)^{F'} \psi_0^0\psi_0^{p+1}\ldots \psi_0^9 \, ,
\ee
in the R sector. Here $(-1)^{F'}$ changes the sign of all the
fermionic oscillators except the zero modes, and
$|s\ra$ and $|s'\ra$ are the Fock space states corresponding to
the vertex operators $V_s$ and $V_{s'}$. 
$\eps$ is a phase factor which is taken to be 1 if $U^2=1$ and
$i$ if $U^2=-1$.
{}From the structure of
\refb{e1} we see that the physical states in the 0-$p$ sector
are in one to one correspondence to the set of $(-1)^F$ even vertex 
operators,  although the actual construction of the vertex
operator involves both $(-1)^F$ even and $(-1)^F$ odd operators.

In the $p$-0 sector the physical vertex operators are taken to be
of the form:
\be \label{e2}
\pmatrix{1 & 0} \otimes V_s - i\eps^*
\pmatrix{0 & 1}\otimes V_{s'}\, ,
\ee
where $V_s$ is a $(-1)^F$ even vertex operator, and
$V_{s'}$ and $V_s$ are again related as in eq.\refb{e3}.
Since $\Omega$ relates the 0-$p$ sector to the $p$-0 sector, in
type I string theory we do not have independent contribution to
the spectrum from the 0-$p$ and the $p$-0 sectors.

\noindent{\bf Rules for computing amplitudes}: Let us now specify
the details of the combinatoric factors associated with a given
amplitude. The most convenient way to specify these rules is
to give a comparison with the amplitude associated with a
D-particle of type IIA string theory. Compared to the
various combinatoric factors in that theory, we have the
following extra factors:
\begin{itemize}
\item For every hole with its boundary
lying on the type IIB D-particle, we
associate a factor of ${1\over \sqrt 2}\times Tr(I)=\sqrt 2$.
\item Around every hole whose boundary lies on a D-particle, 
we only impose periodic boundary
condition on the fermion fields, instead of summing over periodic
and anti-periodic boundary conditions.\footnote{Specifying the
boundary condition on the fermions requires a choice of
coordinate system. We are using the coordinate system in which
the hole corresponds to a circle around the origin of the complex
plane.}
\item For a world sheet boundary with insertion of open string
vertex operators, different segments of the boundary can lie on
different D-branes; hence we cannot specify separate rules for
separate D-branes. The rule to be followed here is that
we should sum over both
periodic and anti-periodic boundary conditions on the fermions.
We can
regard the world sheet diagram with anti-periodic boundary
condition as having a cut associated with the $(-1)^F$ operator ending
on the boundary, If
the cut ends on a segment lying on a $p$-brane for $p\ne 0$ we
use the usual rules of taking trace over the product of Chan
Paton factors. On the other hand {\it if the cut lands on a
D-particle, then we insert an extra Chan Paton matrix $\sigma_2$
on the segment where the cut ends}. Of course, the final answer
should not depend on how we choose the locations of the various
cuts on the world-sheet.
\end{itemize}

This concludes our description of the rules for computing the
spectrum and interactions of the open strings with one or both
ends on the D-particle. Our next task is to check the consistency
of this prescription. For this we first need to check the closure
of the open string operator algebra. In particular, we need to
make sure that in an arbitrary product of various operators, we
do not generate any operator outside the set that
we have included in
the spectrum. We see this as follows:
\begin{enumerate}
\item First we shall check that if we take the product of 
a 0-0 and a 0-$p$ sector vertex
operator, we again get back a vertex operator of the form
\refb{e1}. We shall carry out the analysis in the case where the
0-0 sector vertex operator is in the NS sector. If this
vertex operator is of the form 
$\OO\otimes I$ with $\OO$ being an $(-1)^F$ even operator, then
the operator product has the form:
\be \label{ex1}
\pmatrix{1\cr 0} \otimes V_t + i\eps\pmatrix{0\cr 1} 
\otimes V_{t'}\, ,
\ee
where the states $|t\ra$, $|t'\ra$ are given by:
\be \label{ex2}
|t\ra = \OO |s\ra, \qquad |t'\ra = \OO |s'\ra\, .
\ee
Since $p$ is odd, in the NS (R) sector the operator
$(-1)^{F'}\psi_0^1\ldots\psi_0^p$\break
\noindent ($i(-1)^{F'}\psi_0^0\psi_0^{p+1}\ldots\psi_0^9$) 
commutes with all 
oscillators and hence $\OO$. This shows that $|t\ra$ and $|t'\ra$ are
related in the same way as $|s\ra$ and $|s'\ra$ are related in
eq.\refb{e3}. Thus the resulting vertex operator
has the form of \refb{e1}.

If instead the 0-0 sector vertex operator has the form
$\OO'\otimes \sigma_2$, with $\OO'$ a $(-1)^F$ odd operator, then
the operator product has the form:
\be \label{eyy1}
\pmatrix{1\cr 0} \otimes V_u + i\eps\pmatrix{0\cr 1} 
\otimes V_{u'}\, ,
\ee
where the states $|u\ra$, $|u'\ra$ are given by:
\be \label{eyy2}
|u\ra = \eps\OO' |s'\ra, \qquad |u'\ra = \eps^{-1}\OO' |s\ra\, .
\ee
Again it is easy to verify that $|u\ra$ and $|u'\ra$ are related
in the same way as $|s\ra$ and $|s'\ra$ are in eq.\refb{e3}.
Thus the product has the structure of \refb{e1}.

\item In the same way one can verify that the product of a $p$-0
and 0-0 sector vertex operator has the form of \refb{e2}.

\item Now consider the product of a $p$-0 and 0-$q$ vertex
operator. From \refb{e1} and \refb{e2} we see that this product
has the structure
\be \label{exx1}
V_sV_t+\eps_1\eps_2^*V_{s'}V_{t'}\, ,
\ee
where $\eps_1$ and $\eps_2$ are the $\eps$-factors associated
with the $p$-0 and 0-$q$ sectors respectively.
Here $V_s$ and $V_t$ are $(-1)^F$ even, and $V_{s'}$ and
$V_{t'}$ are $(-1)^F$ odd. Thus the product is a $(-1)^F$ even
operator and represents an allowed vertex operator in the 
$p$-$q$ sector.

\item Finally we consider the product of a 0-$p$ and a $p$-0
vertex operator. 
For simplicity we shall carry out the analysis for the case where
both vertex operators are in the NS sector or both are in the R
sector, but the result is valid even when one is in the NS sector
and the other is in the R sector.
{}From \refb{e1} and \refb{e2} we see that this
has the form:
\ben \label{exx2}
&& \pmatrix{1\cr 0}\pmatrix{1 & 0} \otimes V_s(x) V_t(y)
+ \pmatrix{0\cr 1}\pmatrix{0 & 1} \otimes
V_{s'}(x) V_{t'}(y)\nonumber \\
&& -i \eps^*\pmatrix{1\cr 0}\pmatrix{0 & 1}\otimes
V_s(x) V_{t'}(y)
+ i \eps\pmatrix{0\cr 1}\pmatrix{1 & 0} \otimes
V_{s'}(x) V_{t}(y)\, .
\een
{}From the relation \refb{e3}  between $|s\ra$,
$|s'\ra$, and using the definition of $\eps$,
we can easily verify that
\be \label{exx3}
\la s|\OO|t\ra = \la s'|\OO|t'\ra, \qquad
\la s|\OO|t'\ra = \eps^2\la s'|\OO|t\ra, 
\ee
for all operators $\OO$. This gives the relations:
\be \label{e4}
V_s(x) V_t(y) = V_{s'}(x) V_{t'}(y), \qquad
V_s(x) V_{t'}(y) = \eps^2 V_{s'}(x) V_{t}(y)\, .
\ee
Using eqs.\refb{e4} we can rewrite \refb{exx2} as
\be \label{exx4}
I \otimes V_s(x) V_t(y) + \eps^*
\sigma_2 \otimes V_s(x) V_{t'}(y)\, .
\ee
Both of these are allowed operators in the 0-0 sector. Thus we
see that the product of vertex operators in the 0-$p$ and $p$-0
sectors closes on the allowed vertex operators in the 0-0 sector.
\end{enumerate}

This establishes the closure of the operator algebra in the open
string sector. Next we need to verify the closure of the
open-closed operator algebra. In particular, since the spectrum
of open strings contains both $(-1)^F$ odd and $(-1)^F$ even
states, one might worry if the interaction of such open strings
could produce closed string states carrying odd world-sheet
fermion number. We shall now show that this does not happen. For
this let us consider a disk amplitude with open string vertex
operators inserted at the boundary and a closed string vertex
operator inserted in the interior of the disk. We shall only
consider the cases of NSNS and RR sector closed string vertex
operators. Using the
closure of the open string operator algebra, we can reduce this
to a diagram where only the 0-0 open strings are inserted at the
boundary. In this sector, the $(-1)^F$ even states are accompanied by
the Chan Paton factor $I$ whereas the $(-1)^F$ odd states are
accompanied by the Chan Paton factor $\sigma_2$. 
First let us take
the closed string vertex operator to be in the NS-NS sector so
that the fermions satisfy periodic boundary condition along the
boundary of the disk and hence there is no extra Chan Paton
factor besides those coming from the open string vertex
operators. In this case an
amplitude containing odd number of $(-1)^F$ odd open string
states will have a net Chan factor factor $\sigma_2$ inserted at
the boundary and will vanish upon taking the trace. Hence
the only non-vanishing amplitudes are those with even number of
$(-1)^F$ odd open string states. This, in turn guarantees that
the closed string vertex operator inserted in the interior of the
disk must have the correct GSO projection in order to get a
non-zero amplitude. On the other hand if the closed string vertex
operator is in the RR sector, then there will be an extra Chan
Paton factor of $\sigma_2$ at the boundary, and hence now
non-vanishing trace over the Chan Paton factor will require an
odd number of $(-1)^F$ odd open string
vertex operators. This would indicate that
the closed string vertex operator must be $(-1)^F$ odd. However
note that in the language of type IIA string theory (which we are
using as the reference for comparison) the allowed RR-sector
vertex operators in the type IIB theory are indeed odd under
$(-1)^F$. Thus the interaction rules again guarantee that
only those closed string states which satisfy the correct GSO
projection condition of type IIB string theory are produced in
the scattering of open strings. 

The above analysis also explains the reason for the extra Chan
Paton factor
of $\sigma_2$ that needs to be put in if a cut lands on a segment
of the boundary lying on the D-particle. Let us
consider a 0-$p$ vertex inserted on a boundary, with a cut
passing through the segment lying on the $p$-brane. 
As the cut moves from the 
$p$-brane to the 0-brane, we naturally 
shift from the IIB to the IIA convention for defining
the $(-1)^F$ quantum numbers of spin
fields.\footnote{This is related to the fact that the RR sector
of a boundary state describing a D-particle
can only be defined for the type IIA string
theory, and not for type IIB string theory.} This means that the
definition of $(-1)^F$ odd and $(-1)^F$ even spin fields get
reversed. Since the operator $\epsilon U$ defined in  \refb{e3a},
\refb{e3b} anti-commutes with $(-1)^F$, this shift in the
definition of $(-1)^F$ is generated by multiplying the 0-$p$
vertex operators by $\eps U$. On the other hand, from \refb{e1}
we can easily verify that multiplying a 0-$p$ vertex operator by
$\eps U$ is equivalent to multiplying it by $\sigma_2$. This
shows that as the cut moves from the $p$-brane to the 0-brane, it
is accompanied by an extra Chan Paton factor of $\sigma_2$.

Let us now turn to the consistency of the annulus diagram. In
particular we need to show that the annulus diagram admits a
double interpretation, $-$ as an open string loop diagram and as
a closed string tree diagram. We
shall only discuss the case where both boundaries of the annulus
lie on the D-particle, but the consistency of other diagrams with
one boundary on the 0-brane and the other boundary on a $p$-brane
can also be checked.  Since each boundary gives a 
contribution proportional to $Tr(I)/\sqrt 2=\sqrt 2$, we get a
net factor of 2 compared to the annulus diagram of the type IIA
D-particle. At the same time, we put only periodic boundary
condition on the fermions along the non-trivial cycle, instead of
summing over periodic and anti-periodic boundary conditions. Since
the annulus diagram for the D-particle in type IIA string theory
is known to be given by an integral of
\be \label{ey1}
Tr\Big({1+(-1)^F\over 2} q^{L_0}\Big)\, ,
\ee
we see that the annulus diagram for the D-particle in IIB has the
form of an integral over
\be \label{ey2}
Tr(q^{L_0})\, .
\ee
Here $q$ is a modular parameter to be integrated over.
The absence of the $(-1)^F$ term in \refb{ey2}
is a reflection of the absence
of anti-periodic boundary condition on the fermions. The extra
multiplicative factor of 2 is due to the contribution from the
Chan Paton factors. \refb{ey2} can be clearly
interpreted as the partition function of the open string states
without any GSO projection, and hence can be identified as the
partition function of an open string with both ends on the type
IIB D-particle. 

We can also reinterprete the annulus diagram in the closed string
channel. In the case of type IIA D-particle, the ${1\over
2}Tr(q^{L_0})$ term may be interpreted as the result of interaction
between the D-particles due to NSNS sector
closed string exchange, and the
${1\over 2}Tr\big((-1)^Fq^{L_0}\big)$ term may be interpreted as
the result of interaction due to the exchange of RR sector closed
string exchange. Comparing this with \refb{ey2} we see that in
the present case there is no contribution from the 
exchange of RR sector closed strings, 
whereas the contribution from the NSNS sector exchange
doubles. Thus the D-particle of type IIB does not carry any RR
charge, and is $\sqrt 2$ times heavier than the D-particle of
type IIA. This gives the type IIB D-particle mass to be
\be \label{ey4}
{\sqrt 2\over g} (\alpha')^{-{1\over 2}}\, ,
\ee
where $g$ is the string coupling constant and
$(2\pi\alpha')^{-1}$ is the string tension.  

This extra $\sqrt 2$
factor in the mass can also be seen from the fact that the disk
tadpole for the graviton has an extra factor of $\sqrt 2$ for the
D0-brane of IIB compared to that of IIA, due to the extra
contribution of ${1\over \sqrt 2} Tr(I)$ from the Chan Paton
factor. On the other hand, the disk tadpole for an RR sector
state vanishes in this theory due to periodic bounday condition
on the fermions along the boundary of the disk.

This result can be stated in a different way by constructing the
boundary state for the D-particle of IIB. This has been given in
\cite{BERGAB}. Let us denote the boundary state describing the
D-particle in type IIA by
\be \label{ey5}
{1\over \sqrt 2} (|NSNS\ra + |RR\ra)\, ,
\ee
where $|NSNS\ra$ and $|RR\ra$ denote the contribution from the
NSNS and RR sectors respectively. Then the boundary state
describing the D-particle of type IIB is given by
\be \label{ey6}
|B\ra = |NSNS\ra\, .
\ee
By construction,
the inner product of this boundary state with itself produces the
partition function of the 0-0 string.
But as a further consistency check, one can verify that
the inner product of this boundary state with the boundary state
of a $p$-brane in type IIB correctly reproduces
the partition function of the 0-$p$ string. Furthermore, in type
I string theory we also need to take the inner product of this
with the crosscap boundary state $|C\ra$\cite{POLCAI,CANAYO}. 
This combined with $\la B|B\ra$ can be shown to give the correct
$\Omega$ projected open string partition function in the 0-0
sector.

This finishes our analysis of the consistency checks for the
interaction
rules given at the beginning of this paper. We shall now use
these rules to study some properties of the D-particle. First we
note that even though the D-particle in type IIB is unstable due
to the presence of the tachyonic ground state in the NS sector,
this tachyon is projected out in type I string theory (as
discussed below eq.\refb{en2}). Thus the D-particle is stable in
type I string theory. In order to find its quantum numbers, we
need to find the spectrum of massless fermionic open string
states living on the D-particle and quantize them. Again the
easiest way to proceed is by comparing the present situation to
that of a D-particle of type IIA string theory. The Ramond sector
ground state in the 0-0 sector gives rise to massless fermions as
usual, but due to the absence of GSO projection,
the number of such fermionic states living
on the type IIB D-particle is double of that living on the type
IIA D-particle.  Embedding
this in type I halves the number of fermion zero modes. Thus the
number of fermion zero modes living on the type I D-particle is
{\it equal} to the number of such zero modes living on the type
IIA D-particle. Since the type IIA D-particle is 256-fold
degenerate, we see that the type I D-particle also acquires a
256-fold degeneracy. This corresponds to a long multiplet of the
N=1 supersymmetry in ten dimensions, exactly as expected of a
non-BPS state.

But the type I D-particle has further degeneracy from the 0-9
sector strings. Since the type I vacuum contains 32 nine branes,
we get 32 extra massless fermions from the Ramond sector ground
state of the 32 0-9 strings. These zero modes form a 32
dimensional Clifford algebra, and their quantization gives rise
to a state in the spinor representation of
SO(32).\footnote{Actually we get states in spinor as well as
conjugate spinor representation of the gauge group. However,
there is a residual $Z_2$ gauge symmetry on the world-volume of
the D-particle, as in \cite{POLWIT}, under which the 0-9 strings
are odd, and requiring invariance under this gauge transformation
projects out states in the conjugate spinor representation. I
wish to thank O. Bergman and M. Gaberdiel for discussion on this
point.} Thus we see that the type I D-particle describes a
stable, non-BPS state in the spinor representation of SO(32).
This is precisely the state we have been looking for.

The
next question we shall address is the stability of
the D-particle in the presence of various other branes. First
let us consider the case of two such D-particles on top of each
other. In this case the $\Omega$ projection is no longer
sufficient to project out the tachyonic mode that originates from
an open string whose two ends lie on the two different
D-particles.
Thus this system becomes unstable. This is not surprising, since
a pair of such D-particles has SO(32) quantum number in the
scalar conjugacy class. Thus such a system can roll down to
the vacuum (possibly with other massless particles). The
tachyonic mode simply represents the possibility of such a
transition.

In general,
when we bring the D-particle on top of a D$p$-brane, the NS
sector ground state of the open string with one end on the
D0-brane and the other end on the D$p$-brane corresponds to a
particle with squared mass
\be \label{ez1}
m^2 = -{1\over 2} (1 -{p\over 4}) (\alpha')^{-{1\over 2}}\, ,
\ee
Thus we see that the D0-D1 system has a
tachyonic mode whereas the D0-D5 and D0-D9 system has no
tachyonic mode. The tachyonic instability of the D0-D1 system
again reflects the existence of a lower energy configuration
carrying the same quantum number. To see this let us consider the
case of type I string theory compactified on a circle,
with the D-string wrapped on the circle. In this case the
spinor charge of the D0-brane can be absorbed into the
D-string by putting a $Z_2$ Wilson line along the circle at no
cost of energy\cite{POLWIT,NARA}. 
The tachyonic instability of the D0-D1 system
simply reflects the possibility of rolling down to this lower
energy configuration.

Thus we see that the D-particle of type I described in this paper
not only satisfies the mathematical conditions required for
consistent open-closed interaction, but also satisfies all the
physical properties expected of such a particle.

\bigskip

\noindent{\bf Acknowledgement}: I wish to thank O. Bergman and M.
Gaberdiel for useful communications.

\end{document}